\documentclass[amsmath,amssymb,prl,hyperlink,twocolumn]{revtex4-1}
\usepackage{graphicx}
\usepackage{epstopdf}
\usepackage{epsfig}
\usepackage{soul}
\usepackage[colorlinks=true,citecolor=blue,linkcolor=magenta]{hyperref}
\usepackage[usenames]{color}
\usepackage{amsfonts}

\usepackage{times}

\begin{document}

\title{Quantum State Transfer from a Single Photon to a Distant Quantum-Dot Electron Spin}


\author{Yu He}
\affiliation{Hefei National Laboratory for Physical Sciences at Microscale and Department of Modern Physics, University of Science and Technology of China, Hefei, Anhui 230026, China}
\affiliation{CAS Center for Excellence and Synergetic Innovation Center in Quantum Information and Quantum Physics, University of Science and Technology of China, Hefei, Anhui 230026, China}
\author{Yu-Ming He}
\affiliation{Hefei National Laboratory for Physical Sciences at Microscale and Department of Modern Physics, University of Science and Technology of China, Hefei, Anhui 230026, China}
\affiliation{CAS Center for Excellence and Synergetic Innovation Center in Quantum Information and Quantum Physics, University of Science and Technology of China, Hefei, Anhui 230026, China}
\author{Yu-Jia Wei}
\affiliation{Hefei National Laboratory for Physical Sciences at Microscale and Department of Modern Physics, University of Science and Technology of China, Hefei, Anhui 230026, China}
\affiliation{CAS Center for Excellence and Synergetic Innovation Center in Quantum Information and Quantum Physics, University of Science and Technology of China, Hefei, Anhui 230026, China}
\author{Xiao Jiang}
\affiliation{Hefei National Laboratory for Physical Sciences at Microscale and Department of Modern Physics, University of Science and Technology of China, Hefei, Anhui 230026, China}
\affiliation{CAS Center for Excellence and Synergetic Innovation Center in Quantum Information and Quantum Physics, University of Science and Technology of China, Hefei, Anhui 230026, China}
\author{Kai Chen}
\affiliation{Hefei National Laboratory for Physical Sciences at Microscale and Department of Modern Physics, University of Science and Technology of China, Hefei, Anhui 230026, China}
\affiliation{CAS Center for Excellence and Synergetic Innovation Center in Quantum Information and Quantum Physics, University of Science and Technology of China, Hefei, Anhui 230026, China}
\author{Chao-Yang Lu}
\affiliation{Hefei National Laboratory for Physical Sciences at Microscale and Department of Modern Physics, University of Science and Technology of China, Hefei, Anhui 230026, China}
\affiliation{CAS Center for Excellence and Synergetic Innovation Center in Quantum Information and Quantum Physics, University of Science and Technology of China, Hefei, Anhui 230026, China}
\author{Jian-Wei Pan}
\affiliation{Hefei National Laboratory for Physical Sciences at Microscale and Department of Modern Physics, University of Science and Technology of China, Hefei, Anhui 230026, China}
\affiliation{CAS Center for Excellence and Synergetic Innovation Center in Quantum Information and Quantum Physics, University of Science and Technology of China, Hefei, Anhui 230026, China}

\author{Christian Schneider}
\affiliation{Technische Physik, Physikalisches Instit\"{a}t and Wilhelm Conrad R\"{o}ntgen-Center for Complex Material Systems, Universitat W\"{u}rzburg, Am Hubland, D-97074 W\"{u}zburg, Germany}
\author{Martin Kamp}
\affiliation{Technische Physik, Physikalisches Instit\"{a}t and Wilhelm Conrad R\"{o}ntgen-Center for Complex Material Systems, Universitat W\"{u}rzburg, Am Hubland, D-97074 W\"{u}zburg, Germany}
\author{Sven H\"{o}fling}
\affiliation{Technische Physik, Physikalisches Instit\"{a}t and Wilhelm Conrad R\"{o}ntgen-Center for Complex Material Systems, Universitat W\"{u}rzburg, Am Hubland, D-97074 W\"{u}zburg, Germany}

\date{\vspace{0.3cm}\today}

\begin{abstract}
Quantum state transfer from flying photons to stationary matter qubits is an important element in the realization of quantum networks. Self-assembled semiconductor quantum dots provide a promising solid-state platform hosting both single photon and spin, with an inherent light-matter interface. Here, we develop a method to coherently and actively control the single-photon frequency bins in superposition using electro-optic modulators, and measure the spin-photon entanglement with a fidelity of $0.796\pm0.020$. Further, by Greenberger-Horne-Zeilinger-type state projection on the frequency, path and polarization degrees of freedom of a single photon, we demonstrate quantum state transfer from a single photon to a single electron spin confined in an InGaAs quantum dot, separated by 5 meters. The quantum state mapping from the photon's polarization to the electron's spin is demonstrated along three different axis on the Bloch sphere, with an average fidelity of $78.5\%$.
\end{abstract}


\maketitle

Self-assembled semiconductor quantum dots (QDs) \cite{1QD-review-1,1QD-review-2} have received considerable attention for quantum information processing. They can serve as narrow-linewidth single-photon sources with a near-unity quantum efficiency, high photon indistinguishability, and high extraction efficiency in monolithic microcavities \cite{2QD-photon-2,2QD-photon-3,2QD-photon-4,2QD-photon-5,2QD-photon-6}. Furthermore, QDs have been deterministically charged with single electrons or holes with long spin coherence time \cite{3QD-coherence-1}. The confined spin state has been initialized by optical cooling \cite{4QD-cooling-1,4QD-cooling-2,15Raman-2} and coherently controlled using picosecond laser pulses \cite{5QD-rotation-1,5QD-rotation-2}. The optical selection rules in a singly charged QD provides a high-fidelity quantum entanglement between the electron spin and the emitted photon¡¯s frequency and polarization. Previous demonstrations of QD spin-photon entanglement \cite{7spinphotonentanglement-0} relied on fast photon detectors to resolve the frequency superposition passively \cite{7spinphotonentanglement-1,7spinphotonentanglement-2,7spinphotonentanglement-3,7spinphotonentanglement-4}, and the quantum teleportation from a single photon to a QD spin exploited two-photon interference on a beam splitter which was inherently probabilistic \cite{14gao}.

In this Letter, we develop a new technique for active measurement of single-photon frequency-bin superposition using a phase-locked electro-optic modulator ($p$-EOM). We also demonstrate quantum information transfer \cite{6network} from a single photon to a distant electron spin by Greenberger-Horne-Zeilinger (GHZ) state projection on the frequency, path and polarization degrees of freedom of the single photon. A layout of the experiment is depicted in Fig.~1a. Suppose Alice has a negatively charged single InGaAs QD housed in a 4.2$\,$K bath cryostat. With an external magnetic field of 2.8$\,$T applied in Voigt geometry, the spin ground states $|\downarrow\rangle$, $|\uparrow\rangle$ and one of the trion states $|\downarrow\uparrow\Downarrow\rangle$ form a $\Lambda$ system (see left inset of Fig.~1a). Bob, who is separated by 5 meter from Alice, aims to remotely prepare Alice's QD spin in an arbitrary superposition state which Alice doesn't know.

\begin{figure*}[tb]
\includegraphics[width=0.8\textwidth]{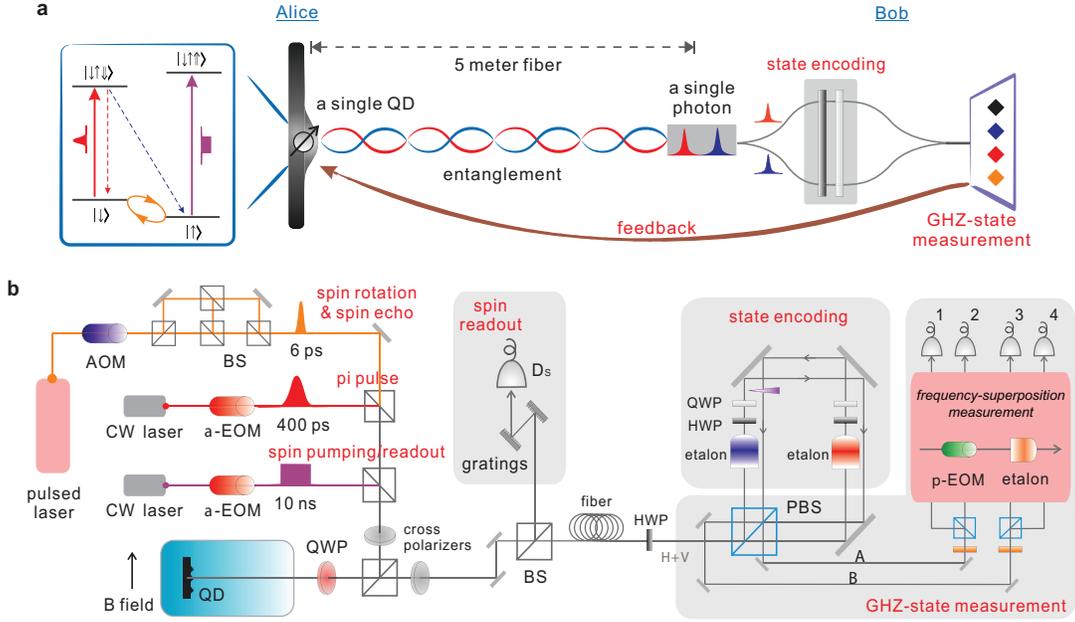}
\caption{Protocol and experiment setup for photon-to-spin state transfer.  \textbf{a}, Alice has a negatively charged QD (see left inset for its energy level under an in-plane magnetic field). Bob, who is at a distant location, aims to prepare the Alice's QD spin in arbitrary superposition state. Alice first generates spin-photon entanglement, and then sends the frequency-encoded photon qubit to Bob. Bob uses a specially designed interferometer (see text for details) to prepare the to-be-teleported state in the photon's polarization. Finally, the polarization, frequency and path degrees of freedom of the photon are measured jointly on four GHZ state basis. By implementing appropriate feedback unitary operations conditioned on the GHZ measurement results, the photon polarization is deterministically transferred to the QD spin. \textbf{b}, Optical arrangement of the experimental setup (see Supplemental Materials for more details). A 10-ns pulse generated by an amplitude electro-optic modulator ($a$-EOM) is used for spin initialization/measurement. A 400-ps pulse is used for deterministic spin-photon entanglement generation. The pulsed laser is modulated by an acousto-optic modulator (AOM) for spin rotation and spin echo pulse sequences. A Sagnac-type interferometer is used to both prepare the to-be-teleported photon polarization state and to perform the GHZ-state measurement. The upper-right inset shows the frequency qubit measurement module, which consists of a phase-locked electro-optic modulator ($p$-EOM) and an etalon. The pink frequency-superposition measurement box contains four frequency qubit measurement modules with one in each optical path.}
\label{fig1}
\end{figure*}

\begin{figure}[tb]
        \includegraphics[width=0.50\textwidth]{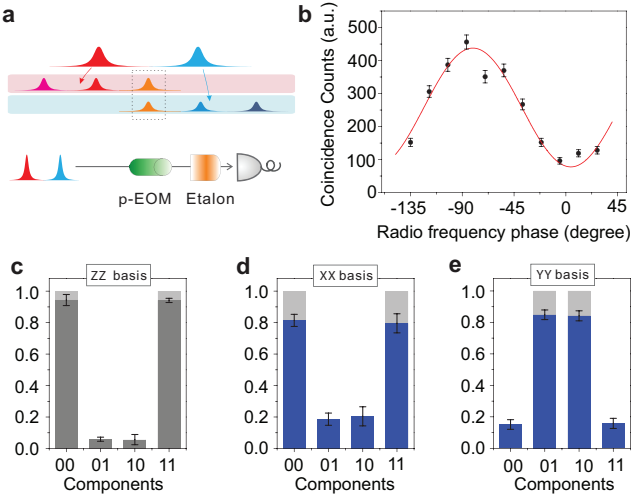}
\caption{Photon frequency qubit measurement and spin-photon entanglement. \textbf{a}, Modulated by a $p$-EOM, the red and blue sidebands of the frequency qubit are transformed into triple peaks, with their relative phase inherited. The overlapped peaks are filtered out with an etalon, where the phase is converted to the field probability amplitude. \textbf{b}, Measured coincidence counts for spin-photon correlation while varying the driving RF field phase delay. \textbf{c-e,} Spin-photon entanglement state normalized coincidence counts on correlated measurement basis. The light gray gap shows difference between ideal and experimental values. All the basis are encoded in the sequence of $|\mathrm{spin}\rangle$$|\mathrm{photon}\rangle.$ For the spin qubit,  $|0\rangle$($|1\rangle$) of Z, X, Y basis (corresponding to the Pauli matrices $\sigma_{z}$, $\sigma_{x}$, $\sigma_{y}$) are encoded as $|\downarrow\rangle$($|\uparrow\rangle$), $(|\downarrow\rangle+|\uparrow\rangle)/{\sqrt2}$ ($(|\downarrow\rangle-|\uparrow\rangle)/{\sqrt2}$), and $(|\downarrow\rangle+i|\uparrow\rangle)/{\sqrt2}$ ($(|\downarrow\rangle-i|\uparrow\rangle)/{\sqrt2}$), respectively. While for the frequency qubit, $|0\rangle$($|1\rangle$) is defined as $|\omega_{red}\rangle$($|\omega_{blue}\rangle$), $(|\omega_{red}\rangle+|\omega_{blue}\rangle)/{\sqrt2}$ ($(|\omega_{red}\rangle-|\omega_{blue}\rangle)/{\sqrt2}$), and  $(|\omega_{red}\rangle+i|\omega_{blue}\rangle)/{\sqrt2}$ ($(|\omega_{red}\rangle-i|\omega_{blue}\rangle)/{\sqrt2}$), for Z, X, and Y basis respectively.}
\label{fig2}
\end{figure}

\begin{figure}[tb]
        \includegraphics[width=0.42\textwidth]{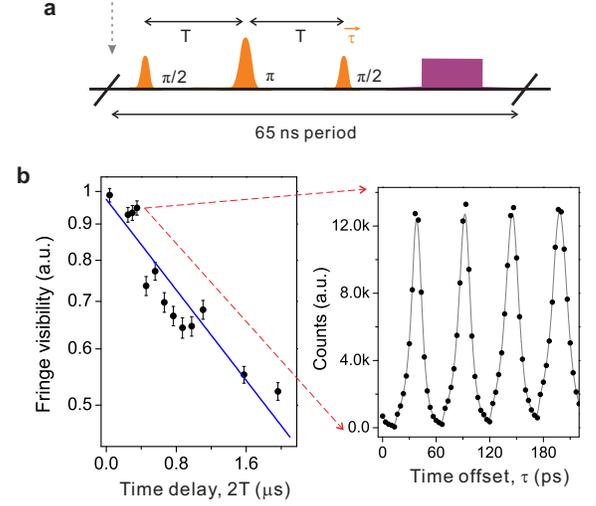}
\caption{Ultrafast optical spin echo for prolonging spin coherence in a single QD. \textbf{a}, Control laser pulse sequence. A first $\pi$/2 pulse generates a spin coherence, followed by a 19-ns time delay during which the spin dephases freely. Next, a $\pi$ rotation is applied, which effectively reserves the direction of spin dephasing. After that, the spin rephases during another 19 ns, at which point another $\pi$/2 pulse is applied to read out the coherence of the spin. \textbf{b}, Measurement of $T_2$ using spin echo. Ramsey interference fringe amplitude on a semilog plot versus time delay of the whole echo pulse sequence, showing a fit to an exponential decay. The inset shows an example of the Ramsey interference fringe at a time delay of 38 ns. The horizontal axis is the delay time of second $\pi$/2 pulse comparing with first $\pi$/2 pulse, zero delay shows where the echo pulse sequence is exactly symmetric.}
\label{fig2}
\end{figure}

\begin{figure*}[tb]
        \includegraphics[width=0.72\textwidth]{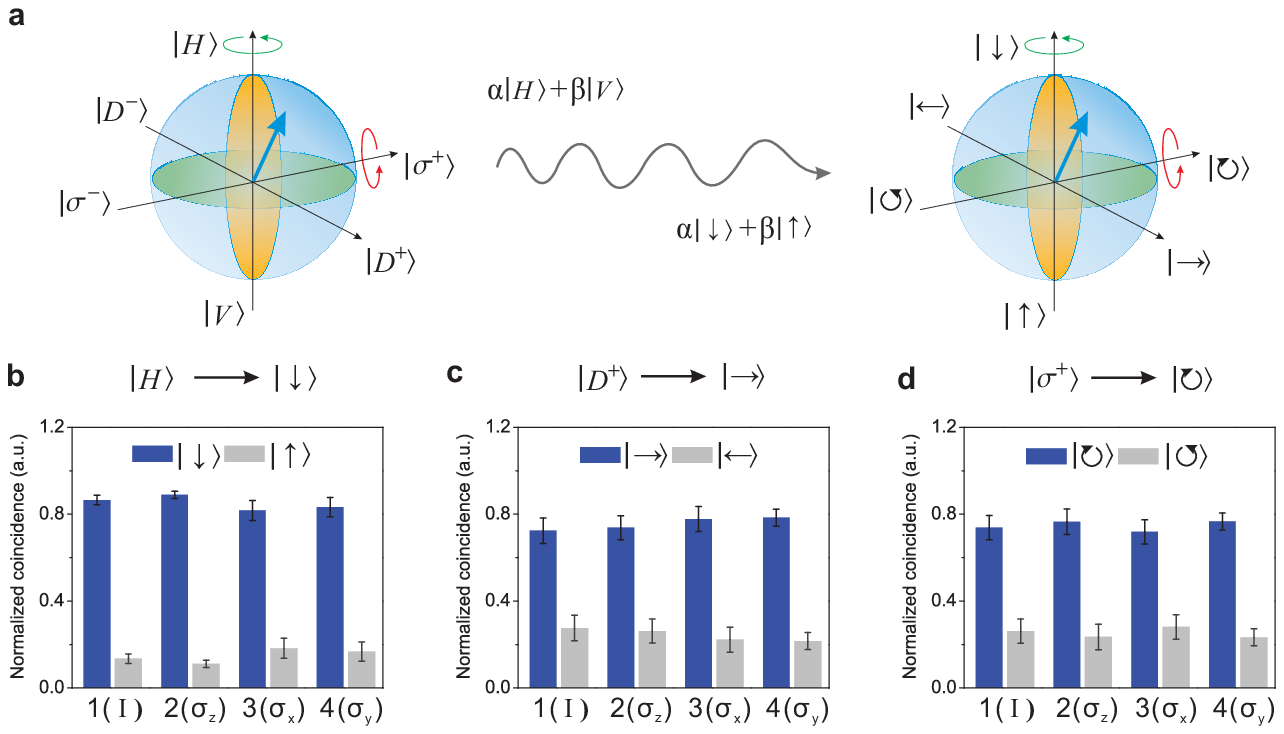}
\caption{Experimental results of quantum state transfer from a single photon to a distant spin.
\textbf{a}, Schematic illustration of photon-to-spin remote state mapping process from photon's Bloch spheres to spin's. \textbf{b}-\textbf{d}, all use blue color to represent ideal outcomes, while gray columns for the undesired coincidence counts on orthogonal basis. The target states are $|H\rangle$ (\textbf{b}), $|D^{+}\rangle$ (\textbf{c}) and $|\sigma^{+}\rangle$ (\textbf{d}), corresponding to correlated spin states of $|\downarrow\rangle$  (\textbf{b}), $|\rightarrow\rangle$  (\textbf{c}) and $|\circlearrowright\rangle$ (\textbf{d}), respectively. For each state, there are four possible GHZ state measurement results such that we incorporate corresponding correction operations ($\sigma_z$, $\sigma_x$, and $\sigma_y$), as shown in the\textit{ x} axis.}
\label{fig4}
\end{figure*}

Firstly, Alice initializes her QD to $|\downarrow\rangle$ by optical cooling, and then near deterministically excite it to $|\downarrow\uparrow\Downarrow\rangle$ by a 400 ps $\pi$-pulse \cite{15Raman-2,15Raman-3, note} (see Fig.~1b). The excited state $|\downarrow\uparrow\Downarrow\rangle$ decays via two possible channels, generating spin-photon entanglement. Two crossed polarizers in the confocal microscope are used to extinguish excitation laser leakage \cite{16RF-2}, meanwhile projecting the photon polarization to be $(|H\rangle-i|V\rangle)/\sqrt{2}$, where $H$($V$) represents horizontal(vertical) polarization. After that, the generated spin-photon entangled state can be written as (see Supplemental Material \cite{suppinfor} for details) \cite{7spinphotonentanglement-1,7spinphotonentanglement-2,7spinphotonentanglement-3}:
\begin{eqnarray}
|\Psi\rangle =\frac{1}{\sqrt2}( | \downarrow \rangle | \omega_{red} \rangle - | \uparrow \rangle | \omega_{blue} \rangle )
\label{entanglement_state}
\end{eqnarray}
where $|\omega_{red}\rangle$ and $|\omega_{blue}\rangle$ are red and blue frequency bins from the two decay channels. This spin-photon entangled state can be directly characterized via active measurement of frequency superposition.

Alice then sends the photon to Bob through a 5-meter optical fiber. Out of the fiber, Bob prepares the photon polarization to be $(|H\rangle+|V\rangle)/\sqrt{2}$. The photon is then split by a polarizing beam splitter (PBS) into two paths, i.e., the $H$ is transmitted ($T$) whereas the $V$ is reflected ($R$), as shown in Fig.~1b. On the two paths, two etalons are placed, and temperature stabilized at the $|\omega_{red}\rangle$ and $|\omega_{blue}\rangle$ frequency bin for the $T$ and $R$ path, respectively. The bandwidth of the etalons are designed to be $\sim\,$1.0 GHz, larger than the single photon's bandwidth ($\sim\,$0.7 GHz) but smaller than the separation of $|\omega_{red}\rangle$ and $|\omega_{blue}\rangle$ ($\sim\,$18.0 GHz).

Hence, the photon's frequency, polarization and path qubits are correlated as: $|\omega_{red} \rangle \rightarrow |\omega_{red} \rangle|H\rangle|T\rangle$ and $|\omega_{blue} \rangle\rightarrow |\omega_{blue} \rangle|V\rangle|R\rangle$. Now, the spin-photon entanglement can be written in a four-qubit Greenberger-Horne-Zeilinger (GHZ) type state:
\begin{eqnarray}
|\Psi'\rangle =\frac{1}{\sqrt2}( | \downarrow \rangle|\omega_{red} \rangle |H\rangle|T\rangle - | \uparrow \rangle |\omega_{blue} \rangle| V\rangle|R\rangle).
\end{eqnarray}

 After that, a half-wave plate (HWP) is inserted in the $R$ path to flip the $V$ polarization to $H$, disentangling the polarization from $|\Psi'\rangle$. The target state to be transferred is encoded in the photon's polarization. Both paths are then placed with a HWP and a quarter-wave plate (QWP) to prepare the polarization in arbitrary superposition: $|\psi\rangle_p=\alpha|H\rangle+\beta|V\rangle$. The composite quantum system can be written as:
\begin{eqnarray}
|\Phi\rangle =\frac{1}{\sqrt2}[|\psi\rangle_p \otimes (| \downarrow \rangle|\omega_{red} \rangle |T\rangle - | \uparrow \rangle |\omega_{blue} \rangle|R\rangle)].
\end{eqnarray}

To achieve photon-to-spin state transfer, in a similar spirit to ref.~\cite{9Bouwmeester97-2} which is a variant of quantum teleportation scheme \cite{8Bennett93}, a crucial step is carrying out joint measurement on the polarization, frequency and path degrees of freedom of the single photon, projecting them onto one of the four GHZ-type states:
\begin{eqnarray}
&&|\xi^{\pm}\rangle=(|H\rangle|\omega_{red} \rangle |T\rangle \pm  | V\rangle|\omega_{blue} \rangle|R\rangle)/\sqrt{2}, \nonumber \\
&&|\chi^{\pm}\rangle= (|H\rangle|\omega_{blue} \rangle |R\rangle\pm  | V\rangle|\omega_{red}\rangle |T\rangle)/\sqrt{2}. \label{eqn4}
\end{eqnarray}

It is remarkable to note that the state $|\Phi\rangle$ can be written in the new basis of these four GHZ-type states,
\begin{eqnarray}
|\Phi\rangle =\frac{1}{2}[ |\xi^{+}\rangle\sigma_{z}+|\xi^{-}\rangle-|\chi^{+}\rangle i\sigma_{y}-|\chi^{-}\rangle\sigma_{x}]\otimes |\psi\rangle_{s}.
\end{eqnarray}
This means that, upon measuring the photon with an equal probability of $1/4$ at one of the four states $|\xi^{+}\rangle$, $|\xi^{-}\rangle$, $|\chi^{+}\rangle$, and $|\chi^{-}\rangle$, and applying simple Pauli corrections $\sigma_{z}$, $I$, $\sigma_{y}$ and $\sigma_{x}$, respectively, the initial state of the photon is transferred to the distant spin, which becomes $|\psi\rangle_{s}=\alpha|\downarrow\rangle+\beta|\uparrow\rangle$.

The above scheme requires a spin-photon entanglement as a quantum resource and two classical bits, which can in principle achieve remote preparation of arbitrary state with 100\% efficiency. A simpler protocol would be to measure the photon state in arbitrary basis and project the spin in a corresponding state. Such protocol is, however, limited to a maximal success probability of 50\% \cite{9Bouwmeester97-2}.

To project and measure the photon in the GHZ-type states, the two paths are combined on the same PBS with a Sagnac-type interferometer. Out of the PBS, the four GHZ states can be separated into two groups: $|\xi^{\pm}\rangle$ exits through output port A, while $|\chi^{\pm}\rangle$ exists through port B, as shown in Fig.~1b. The photon state in port A and B becomes
\begin{eqnarray}
&&|\xi^{\pm}\rangle_A=(|H\rangle|\omega_{red} \rangle \pm  | V\rangle|\omega_{blue} \rangle)/\sqrt{2}, \nonumber \\
&&|\chi^{\pm}\rangle_B= (|H\rangle|\omega_{blue} \rangle \pm  | V\rangle|\omega_{red} \rangle)/\sqrt{2}. \nonumber
\end{eqnarray}
 To further differentiate $|\xi^+\rangle_A$ ($|\chi^{+}\rangle_B$) with  $|\xi^-\rangle_A$ ($|\chi^{-}\rangle_B$), one can analyze the polarization and frequency qubit in the superposition basis $(|H\rangle \pm |V\rangle)/\sqrt{2}$ and $(|\omega_{red}\rangle  \pm |\omega_{blue}\rangle)/\sqrt{2}$. Therefore, the four GHZ-type states correspond to the detection events at four single-photon detectors 1, 2, 3, and 4, as shown in Fig.~1b.

The photon frequency qubit is coherently measured using a $p$-EOM and an etalon.  As shown in Fig.~2a, the $p$-EOM is used to modulate the two frequency bins $|\omega_{red}\rangle$ and $|\omega_{blue}\rangle$ of the photon, where each bin is transformed into three peaks. When the modulation frequency is set at half of the two bins' separation, the blue side band of $|\omega_{red}\rangle$ and the red side band of $|\omega_{blue}\rangle$ overlap with each other, which are then filtered out using an etalon. The intensity of this overlapped bin is proportional to the interference term of $|\omega_{red}\rangle$ and $|\omega_{blue}\rangle$, which thus reflects their relative phase. We control the phase of the driving RF field applied on the $p$-EOM to change the measurement basis of the frequency qubit. The coherent nature of this measurement method can be verified by observing a sinusoidal oscillation by measuring the photon intensity when the state of the spin and photon's polarization is fixed at $(|\downarrow\rangle-|\uparrow\rangle)/\sqrt{2}$ and $(|H\rangle + |V\rangle)/\sqrt{2}$, respectively (see Fig.~2b and Supplemental Materials).

We verify the deterministically generated spin-photon entanglement state in Eq. (1) before performing the state transfer experiment. By replacing the combination of state encoding and GHZ-state measurement modules in Fig.~1b with the frequency qubit measurement module in Fig.~2a, correlation measurements on the spin and frequency qubits can be realized (see Supplemental Material \cite{suppinfor} for setup details). While frequency qubit measurements are achieved by tuning RF field phase as shown in Fig.~2b, spin qubit measurements are accomplished by utilizing rotation pulse and Ramsey precession to transfer the target spin state population to spin $|\uparrow\rangle$. Then read spin $|\uparrow\rangle$ out with a 10-ns pulse where spin-dependent resonance fluorescence photons \cite{19spinreadout,7spinphotonentanglement-1,7spinphotonentanglement-2,7spinphotonentanglement-3,7spinphotonentanglement-4} are registered by a single-photon detector D$_s$, as shown in Fig.~1b. From the histogram of coincidence counts on ZZ basis given by Fig.~2c, we get ZZ basis fidelity $F_{ZZ}=0.942(28)$ , which is mainly degraded by the imperfection of the spin initialization/measurement pulse. Similarly, from the coincidence histograms presented in Fig.~2d and 2e, visibilities $V_{XX}=0.609(51)$ and $V_{YY}=0.690(31)$ are acquired for coherent basis XX and YY, respectively. These visibilities are mainly limited by a spin dephasing time $T_2^{*}=1.7(4)$ ns, where the major dephasing mechanism could be the hyperfine interaction of the electron with the nuclear spins \cite{17nuclear}. Except these aforementioned degrading factors, another common factor is QD re-excitation lead by the 400 ps pulse, which is estimated to degrade fidelity by $6.8 \%$ \cite{note}. Furthermore, based on these three axis correlation measurement results, we obtain a spin-photon entanglement fidelity $F=0.796(20)$, which exceeds the classical limit 0.5 by more than 14 standard deviations.

For the remote state transfer experiment, the electron spin coherence needs to be preserved till the photon propagates about five meters away and is measured. We utilize the optical spin echo technique \cite{18Press} to prolong the spin coherent time. The pulse sequence is shown in Fig.$\,$3a. At the onset of each period, the spin is prepared to the superposition state $(|\downarrow\rangle$+$|\uparrow\rangle)/\sqrt{2}$ by a $\pi/2$ rotation pulse. After a 19-ns spin free precession and dephasing, a $\pi$-pulse reverses the spin precession direction and thus the spin rephases for another 19-ns symmetry evolution. We extract the visibilities of Ramsey interference fringes at different delay time, and obtain the spin decoherence time $T_{2}=2.7\pm0.3$ $\mu$s (see Fig.$\,$3b), which is prolonged for about three orders of magnitudes compared to $T{_2}^*$ and becomes sufficient for our experiment.

To test our scheme works for arbitrary spin superposition states, we prepare three mutually unbiased states along  three different axis on the Bloch sphere. The aim is a faithful state mapping at a distance from the photon's polarization to the spin (see also Fig.$\,$4a):
\begin{eqnarray}
|H\rangle & \Rightarrow & |\downarrow\rangle , \nonumber \\
|D^+\rangle=(|H\rangle + |V\rangle)/\sqrt{2} & \Rightarrow & (|\downarrow\rangle+|\uparrow\rangle)/\sqrt{2} =|\rightarrow\rangle, \nonumber \\
|\sigma^+\rangle=(|H\rangle + i|V\rangle)/\sqrt{2} & \Rightarrow & (|\downarrow\rangle+i|\uparrow\rangle)/\sqrt{2}=|\circlearrowright\rangle. \nonumber
\end{eqnarray}
To evaluate the performance, we measure the state fidelity, i.e., the overlap of the transferred spin state and the ideal one.  The transfer fidelity can be deduced from the coincidence counts of photon detection (detectors 1, 2, 3, or 4) and spin detection (D$_s$). The intrinsic randomness of the outcome of the projected four GHZ-type states is compensated by unitary operations in data post-processing (see Supplemental Materials) depending on the outcome of photon measurement.

Figure~3b-d present the experimental data for input states $|H\rangle$,  $|D^+\rangle$, and $|\sigma^+\rangle$, respectively. The blue bars show the normalized events where the electron spin is measured to be in the correct transferred states, while the gray bars are when the spin ends up in the orthogonal states. We test and average over all the possible four outcomes of the GHZ-type projection (see Eqn.~\ref{eqn4}). From these data, we calculate the fidelities are $\mathcal{F}_{|H\rangle}=0.851\pm0.017$, $\mathcal{F}_{|D^{+}\rangle}=0.756\pm0.027$ and $\mathcal{F}_{|\sigma^{+}\rangle}=0.747\pm0.027$.

We have demonstrated a new protocol of quantum state transfer from a single photon to a single solid-state spin, as a way to remotely prepare single electron spin in arbitrary superposition state. Although the protocol can in principle work deterministically as in Ref.$\,$\cite{9Bouwmeester97-2}, the experimental realization still suffers from various loss, including photon extraction ($\sim$$8\%$), detection ($\sim$$20\%$), single-mode fiber coupling ($\sim$$40\%$), cross polarization ($\sim$$50\%$), wave plates and mirrors ($\sim$$36\%$), and frequency selection loss in the $p$-EOMs ($\sim$$30\%$). The efficiency can be enhanced in the future by, for instance, embedding the QD inside a micropillar cavity \cite{2QD-photon-3,2QD-photon-5,2QD-photon-6,micropillar-11}. The loss only decreased the photon-to-spin state transfer success probability, however, doesn't affect the spin state fidelity. Heralded upon the detection of a single photon after the GHZ-type projection, the distant spin states are demonstrated to be prepared in an arbitrary superposition state with a high fidelity. We expect our results can add a useful toolbox to the investigations of solid-state quantum networks \cite{spin-spin}.

\vspace{0.1cm}
\noindent \textit{Acknowledgement}: We acknowledge S.~K. Gorman for his helpful suggestions. This work was supported by the National Natural Science Foundation of China, the Chinese Academy of Sciences, and the National Fundamental Research Program. We acknowledge financial support by the State of Bavaria and the German Ministry of Education and Research (BMBF) within the projects Q.com-H and the Chist-era project SSQN.

\end{document}